\documentclass[runningheads]{llncs}

\usepackage[T1]{fontenc}
\def\doi#1{\href{https://doi.org/\detokenize{#1}}{\url{https://doi.org/\detokenize{#1}}}}
\usepackage{graphicx}
\usepackage{listings}
\usepackage{algorithm}
\usepackage[noend]{algpseudocode}
\algnewcommand\algorithmicforeach{\textbf{for each}}
\algdef{S}[FOR]{ForEach}[1]{\algorithmicforeach\ #1\ \algorithmicdo}

\usepackage[caption=false]{subfig}

\begin{document}
\title{Towards Schema Inference for Data Lakes}
%
%


\author{Nour Alhammad \and Alex Bogatu \and Norman W Paton}

\institute{Department of Computer Science, \\
University of Manchester, Manchester M13 9PL, UK \\
\email{norman.paton@manchester.ac.uk}}

%
\authorrunning{N. Alhammad et al.}
%
%
\maketitle              
\begin{abstract}
A data lake is a repository of data with potential for future analysis.  
However, both discovering what data is in a data lake and exploring related data sets can take significant effort, as a data lake can contain an intimidating amount of heterogeneous data.  In this paper, we propose the use of schema inference to support the interpretation of the data in the data lake.  If a data lake is to support a schema-on-read paradigm, understanding the existing schema of relevant portions of the data lake seems like a prerequisite.  In this paper, we make use of approximate indexes that can be used for data discovery to inform the inference of a schema for a data lake, consisting of entity types and the relationships between them.  The specific approach identifies candidate entity types by clustering similar data sets from the data lake, and then relationships between data sets in different clusters are used to inform the identification of relationships between the entity types.  The approach is evaluated using real-world data repositories, to identify where the proposal is effective, and to inform the identification of areas for further work.
\keywords{data lake \and data discovery \and schema inference}
\end{abstract}

\section{Introduction}

Data lakes are repositories of data that have been subject to minimal curation, so the data therein can be considered to be largely in its original form~\cite{DBLP:journals/corr/abs-2106-09592}.  Data lakes are created to provide ready access to data that has the potential to be used for analyses.  However, the speculative nature of the data collection and the minimal curation means that data lakes have to be associated with tools and techniques to support data discovery and exploration, for example to populate meta-models that capture their key features~\cite{DBLP:conf/dawak/EichlerGGSM20}.

Data discovery in data lakes can be supported by dataset search~\cite{DBLP:journals/vldb/ChapmanSKKIKG20}. 
However, having identified data sets of interest, there may then be benefit in exploring the data sets, for example by constructing a knowledge graph that builds on the relationships between individual data sets~\cite{DBLP:conf/icde/FernandezMQEIMO18} or by building bespoke data structures for navigation~\cite{DBLP:conf/sigmod/NargesianPZBM20}. This raises the question as to the nature of the most suitable structure for supporting the comprehension and exploration of the data in a data lake.  

In this paper, we propose the use of {\it schema inference} for data lakes, where schema inference seeks to identify entities and relationships that recur within the data sets in the data lake.  The hope in doing so is that an inferred schema, which may be considered to be the conceptual schema of the data lake, can be useful for summarising the contents of the data lake, and for providing the context for the result of a search or discovery task.

The contributions of this paper are as follows:
\begin{enumerate}
	\item The identification of the potential for schema inference, as a complement to existing proposals for discovering and exploring data in data lakes.
	\item The provision of a technique for inferring schemas for data lakes, building on approximate indexing: (i) to provide evidence to inform clustering for identifying candidate entity types, and (ii) to identify relationships between sources that inform the inference of relationships between entity types.
	\item The evaluation of the resulting proposal using real-world data repositories, to identify where the proposal is effective, and to inform the identification of areas for further work.
\end{enumerate}


\section{Related Work}
\label{sec:related}

In this section, we consider related work in two areas of relevance to the current proposal, {\it schema inference} and {\it data lakes}.

\noindent
{\bf Schema inference} is the process of inferring a structure that can be used to characterise heterogeneous individuals.
Schema inference has been applied to different data models for different purposes.  For example, for {\it tabular data}, schema inference has been used to extract tabular data from spreadsheets~\cite{DBLP:conf/das/KociTL018} and text files~\cite{DBLP:journals/pvldb/Christodoulakis20}, using a variety of techniques, including learning to identify different features of tables~\cite{DBLP:conf/edbt/0001VN21} 
and search to explore the space of alternative extractions~\cite{DBLP:conf/icdar/KociT0L19}. 

For {\it graph data}, there are examples of both inference of representations that fully capture the structures of the data in the sources (e.g., \cite{DBLP:journals/vldb/GoasdoueGM20}) and of approaches that expect to operate with highly heterogemeous data, as in the semantic web, where the aim is to identify the recurring features (e.g., \cite{DBLP:conf/er/Kellou-MenouerK15,DBLP:journals/tlsdkcs/ChristodoulouPF15}). 

For {\it hierarchical data}, the main aim has been to infer a suitable schema definition, for example in XML Schema (e.g., \cite{DBLP:conf/vldb/BexNV07}) 
or JSON Schema (e.g., \cite{DBLP:conf/edbt/BaaziziLCGS17}), 
from a document containing instance data.  In such a setting, there is often a trade-off between generating a strict but large schema definition, and a more compact but relaxed definition.

In this paper, schema inference involves a two-phase approach, with clustering followed by relationship inference. Such an approach has been adopted before (e.g., \cite{DBLP:conf/er/Kellou-MenouerK15,DBLP:journals/tlsdkcs/ChristodoulouPF15}). As such, this paper does not provide a radically different approach to schema inference from some prior work, but it does work in a different setting.  For example, unlike \cite{DBLP:conf/er/Kellou-MenouerK15,DBLP:journals/tlsdkcs/ChristodoulouPF15}, we cannot depend on shared vocabularies to inform the identification of entity types, or the definition of explicit links within sources to inform the creation of relationships.


\noindent
{\bf Data Lakes} have given rise to research on a variety of topics; here we review
{\it dataset discovery} and {\it dataset navigation}.  

{\it Dataset discovery} is relevant to this paper because, like schema inference, it aims to support users in becoming familiar with the contents of a data lake.  
We now outline a few recent proposals. Given a target data set, Table Union Search~\cite{Nargesian-18} looks for the tables that are union compatible with the target, building on Locality Sensitive Hashing (LSH) indexes and a probabilistic model to select from different measures of similarity.  D3L~\cite{Bogatu-20} also builds on LSH indexes to match a target, but combines different similarity measures to provide an overall similarity score, and considers how joins of source tables may be able to populate the target more effectively than individual tables. 
These results complement ours, as dataset discovery provides a starting point for data exploration that could benefit from schema inference.  

{\it Dataset navigation} is relevant to this paper because, like schema inference, it identifies relationships between data sets in the data lake, and uses these to inform data access and interpretation.  In Seeping Semantics~\cite{DBLP:conf/icde/FernandezMQEIMO18}, building on indexes that use word embeddings, a graph is created that captures the relationships between data sets. An alternative approach develops a structure specifically for navigating between data sets~\cite{DBLP:conf/sigmod/NargesianPZBM20}, where the goal is to support a small number of steps to identify relevant groups of sources.  Both of these approaches have similar goals to, but complement, the use of schema inference to produce a conceptual model of the concepts in the data lake.

\section{Technical Context}
\label{sec:context}

In this section, we describe two existing technical results on which we build, namely {\it approximate indexing} and clustering using {\it DBSCAN}.

\subsection{Approximate Indexing}
\label{sec:indexing}

Approximate indexing is used in this paper for all of: (i) estimating the similarity of data sets from the data lake; (ii) identifying matches between data set attributes; and (iii) inferring foreign key relationships between source data sets.  

The indexes used to inform schema inference are based on those used for dataset discovery in D3L~\cite{Bogatu-20}, for which the source is available\footnote{https://github.com/alex-bogatu/d3l}.  These indexes use Locality Sensitive Hashing (LSH)~\cite{DBLP:conf/stoc/IndykM98}, and in particular LSH Forest~\cite{DBLP:conf/www/BawaCG05}.  LSH requires hash functions that provide a high probability of placing similar items in the same bucket; different hash functions give rise to indexes with different notions of similarity. 

In D3L, there are indexes for: 
{\it Name:} measuring similarity between $q$-grams of attribute names;
{\it Value:} measuring similarity between the tokens in attribute values;
{\it Format:} measuring similarity between the regular expressions describing attribute values;
{\it Embedding:} measuring similarity between the word embeddings of attribute values; and
{\it Distribution:} measuring the similarity between value distributions for numerical attributes.

The results for the indexes of individual attributes can be combined to derive the similarity of the table as a whole.  The Euclidean distance of these column similarity measures is then used as an overall measure of the similarity of a pair of tables.

\subsection{Clustering}
\label{sec:clustering}

In this paper, we use clustering to group together data sets that are candidates to be represented by the same real-world entity type.  
The clustering algorithm that is being used is DBSCAN~\cite{DBLP:conf/kdd/EsterKSX96}, which was chosen because it doesn't need to know the number of clusters to produce up-front, can produce clusters of arbitrary shape, has few configuration parameters, and scales to accommodate large numbers of data points, which in our case are data sources. The average case performance for {\it n} data points is {\it O(n log n)}.


\section{Identifying Entity Types}
\label{sec:entity}

\begin{figure}[tb]
\centering
\includegraphics[width=0.7\textwidth]{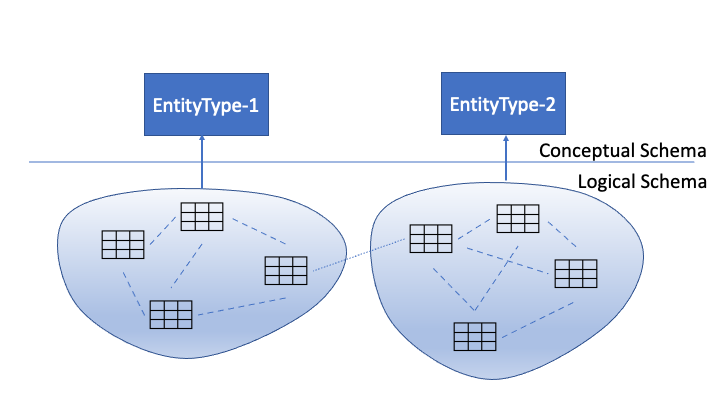}%
\caption{Entity types and clusters.}
\label{fig:clustering}
\end{figure}

In this section, we describe how the clustering algorithm from Section \ref{sec:clustering} has been applied to schema inference for data lakes, using the approximate indexing techniques from Section \ref{sec:indexing}. 
Figure \ref{fig:clustering} illustrates the basic idea. The starting point is a collection of tables that represent the {\it logical schema} of the data lake.  Our aim is to infer the {\it conceptual schema} of the data lake, in the form of a collection of entity types and the relationships between them.  In the figure, the roughly circular shapes represent clusters.  The dashed lines represent similarity relationships between tables that are close enough for DBSCAN to place the associated tables in the same cluster.  The dotted line represents the fact that there are tables that have similarity relationships that are not sufficiently strong to lead to the tables being placed in the same cluster.  All similarity relationships are obtained from the D3L indexes. Each cluster produced by DBSCAN is considered to represent an entity type.


Algorithm \ref{alg:cluster} shows the basic steps.  In line 2, the distance matrix that is used for clustering is populated with maximum distance values, so that closer distances can be assigned for all tables that are found to be similar to each other through index lookup.  Then in lines 3-6, we consider similarity for each of the tables in turn.  For each table we use the D3L similarity index (line 4) to identify its neighbouring tables.  For each of these neighbours (lines 5-6), we update the distance matrix with the measure of their distance from the index.  This approach can be expected to be much more efficient than any all-against-all approach to populating the distance matrix, as related tables are accessed directly by way of the index.  The use of indexes to reduce the comparison space may be considered analogous to index based blocking in entity resolution~\cite{DBLP:journals/csur/PapadakisSTP20}, though we do not follow up with a more detailed comparison step.  For clustering entity types, we are less likely to be dealing with the sort of marginal similarity decisions that can be important in entity resolution.

\begin{algorithm}[tb]
\footnotesize
\caption{Create clusters of tables for given configuration parameters.}\label{alg:cluster}
\begin{algorithmic}[1]
\Function{ClusterTables}{index, k, tables, eps, min\_points}
\State Initialise distance matrix D to max for all positions
\ForEach{$t1 \in tables$}
	\State $neighbours \gets index.lookup(t, k)$
	\ForEach{$(t2,distance) \in neighbours$}
	 	\State Set distance matrix D for $t1$ to $t2$ to $distance$
	\EndFor
\EndFor
\State \Return $DBSCAN(D, k, eps, min\_points)$
\EndFunction
\end{algorithmic}
\end{algorithm}

Algorithm \ref{alg:cluster} assumes that suitable values are available for the two main configuration parameters of DBSCAN: {\it eps}, the maximum distance between two points for them to be considered to be neigbours; and {\it min\_points}, the minimum number of points in a neighbourhood for a point to be considered as a core point, where a core point is on the inside of a cluster rather than on the border.  To automate the selection of values for these parameters, we use a grid search, and select the values with the highest silhouette coefficient~\cite{ROUSSEEUW198753}, which is based on the tightness of elements within clusters relative to the separation of elements from different clusters.  As a result, different experiments identify different values for these parameters.
Algorithm \ref{alg:cluster} also assumes that a suitable value can be provided for $k$, the maximum size of the neighbourhood.  Here some local knowledge on the diversity of the data sets in the data lake may be useful; in practice we can conservatively set $k$ to the number of tables in the data lake, and typcially the index lookup will return many fewer than $k$ results.

\begin{figure}[tb]
\centering
\includegraphics[width=0.7\textwidth]{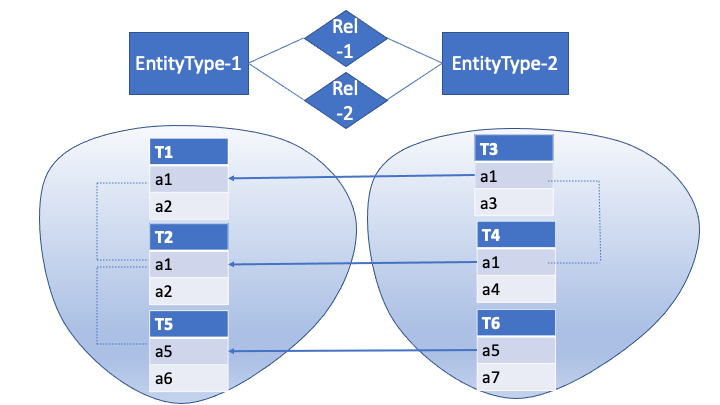}%
\caption{Inferring relationships.}
\label{fig:relationship-details}
\end{figure}

\section{Identifying Relationships}
\label{sec:relationships}

The inference of relationships between candidate entity types involves identifying relationships between the data sources that are associated with those types, and establishing when the presence of such relationships implies the presence of a plausible relationship between the entity types.




Figure \ref{fig:relationship-details} illustrates the approach, where the solid arrows represent foreign key relationships and the dotted lines represent attributes that match (i.e., that are considered by schema matching to represent equivalent attributes).  Figure \ref{fig:relationship-details} depicts two clusters, one containing the tables {\it T1}, {\it T2} and {\it T5}, and the other containing the tables {\it T3}, {\it T4} and {\it T6}.  There are three inferred foreign key relationships between the tables in the cluster, specifically from {\it T3-a1} to {\it T1-a1}, from {\it T4-a1} to {\it T2-a1} and from {\it T6-a5} to {\it T5-a5}.  The question for relationship inference is {\it which of these foreign key relationships represent the same relationships at the conceptual level between the two entity types that represent the cluster}.  The proposal is that they should represent the same relationship when the attributes within the clusters match (depicted by the dotted lines in the figure).  Where the attributes match, they are considered to be equivalent, and thus the associated foreign keys are taken to represent the same notion.  

This automatic grouping of foreign keys to create conceptual relationships is not sure to yield correct results, as the two foreign keys could represent different application semantics.  For example, assume that {\it T1} and {\it T2} both represent cars, with the {\it a1} attributes representing the model. Further assume that {\it T3} and {\it T4} represent people, and the foreign keys from the {\it a1} attributes represent cars associated with those people.  The {\it T3-a1} and {\it T4-a1} attributes of {\it T3} and {\it T4} have matched, so they should represent the same notion. However, if one represents an {\it owns} relationship for a current car and another represents an {\it owned} relationship for a former car, then these relationships will be considered to be the same by the algorithm presented here. However, this reflects the fact that all automated annotation techniques for data lakes are best-effort, which of course also applies to the clustering for identifying types.

To support this, we use Algorithm \ref{alg:relationships}.  This in turn uses additional information that can be obtained using D3L indexes:
\begin{itemize}
\item $Match(Table_i, Attribute_i, Table_j, Attribute_j)$: The $Match$ relationship between $Attribute_i$ of $Table_i$ and $Attribute_j$ of $Table_j$ postulates that the two attributes represent the same notion in the application, based on their similarity, using the attribute-level indexes of D3L.
\item $ForeignKey(Table_i, Attribute_i, Table_j, Attribute_j)$: The $ForeignKey$ relationship between $Attribute_i$ of $Table_i$ and $Attribute_j$ of $Table_j$ postulates that the two attributes are joinable.  The condition for this is that (at least) one of the attributes is a key, and there is a high level of overlap in the {\it Value} index.
\end{itemize}

We also assume that the function $ConceptualType(Table)$ $\rightarrow ConceptualType$, given a source $Table$ returns the identifier of the $ConceptualType$ (cluster) of which it is part.

\begin{algorithm}[tb]
\footnotesize
\caption{Infer relationships between entity types from the relationships between their members.}\label{alg:relationships}
\begin{algorithmic}[1]
\Function{InferRelationships}{clusters,matches,foreign\_keys}
\State $groups \gets \{\;\}$
\ForEach{$(s_i, a_i, s_j, a_j) \in foreign\_keys$}
	\If {$groups = \{\;\}$}
		\State $groups = groups \cup \{\{(s_i, a_i, s_j, a_j)\}\}$
	\Else
		\State added = false
		\ForEach {$g \in groups$}
			\State $MatchCount = 0$
			\ForEach {$(s_i', a_i', s_j', a_j') \in g$}
				\If {($Matches(s_i,a_i,s_i',a_i') \wedge 
					Matches(s_j,a_j,s_j',a_j') \wedge
					EntityType(s_i,clusters) = EntityType(s_i',clusters) \wedge
					EntityType(s_j,clusters) = EntityType(s_j',clusters)$}
					\State MatchCount = MatchCount + 1;
				\EndIf
			\EndFor
			\If {$MatchCount = |g|$}
				\State $g.add((s_i, a_i, s_j, a_j))$
				\State added = true
				\State $break$
			\EndIf
		\EndFor
		\If {$not \; added$}
			\State $groups = groups \cup \{\{(s_i, a_i, s_j, a_j)\}\}$
		\EndIf
	\EndIf
\EndFor
\State \Return $groups$
\EndFunction
\end{algorithmic}
\end{algorithm}

Algorithm \ref{alg:relationships} defines \textsc{InferRelationships}, which takes as input the clusters produced by Algorithm \ref{alg:cluster}, a set of $matches$ and a set of $foreign\_keys$, and returns a set of $groups$. Each group is a set of relationships from $foreign\_keys$ that are postulated to represent a single relationship in the conceptual schema.

The algorithm iterates over each of the $foreign\_keys$ identified between the source tables, to identify if the foreign key should join an existing group of $foreign\_keys$ or should start a new group.  If there are currently no existing groups, then a new group is created containing the current foreign key (lines 4--5).  Where there are existing groups, the current foreign key is considered for membership of each of these groups in turn (lines 7-16).  The current foreign key $(s_i, a_i, s_j, a_j)$ is added to an existing group only if (line 11): (i) $a_i$ matches with the primary key of every foreign key relationship in the group; (ii)  $a_j$ matches the referencing attribute of every foreign key relationship in already in the group; and (iii) the foreign key represents a relationship between the same two clusters as the other members of the group.


Consider the example in Figure \ref{fig:relationship-details}. The relationships between the tables give rise to two relationships at the conceptual level.  One of these relationships, say {\it Rel-1} involves the group of foreign keys  $\{(T1, a1, T3, a1),(T2, a1, T4, a1)\}$; these foreign keys are brought together in the same group because of the matches of {\it T1.a1} with {\it T2.a1} and {\it T3.a1} with {\it T4.a1}. However, the foreign key $(T5, a1, T6, a5)$ gives rise to a second (unary) group because there is no match connecting {\it T6.a5} to {\it T3.a1} and {\it T4.a1}.  The lack of this match is used to conclude that the attribute {\it T6.a5}, and thus the relationship in which it participates, has a different semantics from the attributes (and thus the relationships) of {\it T3.a1} and {\it T4.a1}.

\section{Evaluation}
\label{sec:evaluation}

For the experimental evaluation, we first walk through the experimental setup and the metrics used with a small example; this example allows fairly detailed coverage of the behaviour of the approach that is not practical with real-world cases.  We then apply the approach to open government data and web tables.  Results are in Table \ref{tab:clustering}.

\subsection{Experimental Setup and Walkthrough}

The experiments have been carried out in Python, using the {\it scikit-learn} implementation of {\it DBSCAN}, on an iMac (3.4Ghz Quad Core Intel Core i5, 16Gb Memory). The quality of clustering results is reported using the {\it rand score}, which assesses a clustering as the number of individuals that agree divided by the total number of individuals.  In practice, where a set of elements {\it S} is partitioned into two partitions which consist of sets of custers, {\it X} which is computed and {\it Y} which is the ground truth, this involves the following terms\footnote{https://en.wikipedia.org/wiki/Rand\_index}:
\begin{itemize}
	\item {\it a,} the number of pairs of elements in the same cluster in {\it X} and in the same cluster in {\it Y}.
	\item {\it b,} the number of pairs of elements in different clusters in {\it X} and in different clusters in {\it Y}.
	\item {\it c,} the number of pairs of elements in the same clusters in {\it X} and in different clusters in {\it Y}.
	\item {\it d,} the number of pairs of elements in different clusters in {\it X} and in the same clusters in {\it Y}.
\end{itemize}
Then $rand\_score = \frac{a+b}{a+b+c+d}$.

To evaluate relationships, we use a variant of the rand score.  Relationships between clusters are collections of foreign keys between the tables in the clusters.  These collections of foreign keys can themselves be seen as clusters.  However, the rand score is designed to work on partitions of a fixed set of values, and the foreign keys in a solution are not sure to be drawn from a fixed set: foreign key inference may miss some valid foreign keys (false negatives -- {\it FN}) and may infer some invalid relationships between tables (false positives -- {\it FP}).  As a result, we use a variant of the rand score that includes the following term:
\begin{itemize}
	\item {\it e}, the number of items in $X$ that do not appear in $Y$, or {\it vice versa}, $|FP| + |FN|$.
\end{itemize}
Then the modified rand score for relationships, $rand\_score_R = \frac{a+b}{a+b+c+d+e}$.

For the test dataset, there are {\it 5} tables, {\it 3} relating to {\it people} and {\it 2} relating to movies. As such, there are {\it 2} clusters in the Ground Truth.  However, although DBSCAN produces 2 clusters, the {\it people} cluster contains only {\it 2} of the {\it 3} tables about people. This is because, two of the tables about people have overlapping attribute names and similar values, and the third table about people has disjoint attribute names and largely distinct values.  As such, there is little evidence that this third table should be clustered with the first two.  The  $rand\_score = \frac{a+b}{a+b+c+d}$ thus has $a=2$, $b=6$, $c=2$ and $d=0$, so the overall rand score is $0.8$.

In terms of the relationships, in the ground truth, the cluster containing people is involved in two relationships with the cluster containing films, a {\it likes-movie} relationship and a {\it favourite-movie} relationship.  The two correctly clustered {\it people} tables each implement one of these relationships. The incorrectly clustered {\it people} table has a {\it favourite-movie} attribute, but it is not a foreign key for either of the movie tables.  The $rand\_score_R = \frac{a+b}{a+b+c+d+e}$ thus has $a=2$, $b=4$, $c=0$, $d=0$ and $e=2$, where the $e$ represents the fact that the expected foreign keys between the unclustered person table and the movie tables are missing. As a result, $rand\_score_R = 0.75$.

\begin{table}[t]
\begin{center}
\small
\begin{tabular}{|l|r|r|r|r|}
\hline
Dataset 	       		& Test & Open: NHS & Web: City & Web: Culture \\ 
\hline
Number of Tables       		&  5   & 19   & 50   & 54  \\
Number of Clusters     		&  2   & 2    & 9    & 3 \\
Number of GT Clusters  		&  2   & 3    & 4    & 4 \\
Cluster Rand Score             	& 0.8  & 0.70 & 0.75 & 0.62 \\
Relationships Rand Score        & 0.75 & 0.82 & N/A  & N/A \\
\hline
Build Indexes (s)		& 0.38    & 92.16  & 2.96    & 2.31 \\
Build Distance Matrix (s) 	& 0.19    & 92.57  & 9.24    & 8.26 \\
Cluster (s) 			& 0.0054  & 0.0053 & 0.0057  & 0.0055 \\
Data Profiling (s)		& 0.10    & 8.04   & 1.81    & 1.91 \\	
Infer Relationships (s) 	& 0.00032 & 0.0014 & 0.00033 & 0.0026 \\
\hline
\end{tabular}
\normalsize
\end{center}
\caption{Clustering Results}
\label{tab:clustering}
\end{table}

\subsection{Open Government Data}

The open government data is from the UK National Health Service (NHS), and provides information about the organisations that provide different health services\footnote{https://www.nhs.uk/about-us/nhs-website-datasets/}. In the experiments, these involve {\it 19} data sets, covering different types of organisation (e.g., dentists, pharmacies), opening times, services provided, etc.  The data sets are all produced by the same publisher, and thus are reasonably clean and consistent.  

The ground truth contains three clusters, for organisations, for opening times and for services. The approach produces the clusters for organisations and opening times correctly, but has failed to produce the cluster for services.  This is because the services provided by the different types of organisations are essentially disjoint, and thus the extents of the different services tables have little in common.  However, the overall rand score is quite high, at {\it 0.70}, reflecting the fact that no tables have been incorrectly placed together in clusters.

There are ground truth relationships between organisations, their opening times, and the servives they provide. The relationship between the organisations and their opening times is correctly identified, but the fact that the services are not clustered means that the relationships between organisations and services have not been correctly grouped, giving rise to a relationship rand score of $0.82$.

\subsection{Web Tables}

The web table data is originally from the Web Data Commons corpus, but more specifically is from the T2Dv2 Gold Standard for matching web tables to DBpedia\footnote{http://webdatacommons.org/webtables/goldstandardV2.html}. Of particular relevance to this paper, the gold standard includes manually produced correspondences from web tables to concepts in the DBpedia ontology~\cite{DBLP:conf/edbt/RitzeB17}.  These correspondences, though not specifically designed for use with schema inference, are used as an independently produced ground truth for our experiments; it is assumed that the tables associated with the same concept should cluster together.  We use sub-domains from T2Dv2, involving cities and related concepts (specifically, airports, universities and museums), and using various cultural concepts (including films, musical works and songs).

In relation to the cities data, more clusters are inferred (9) than appear in the ground truth (4).  This is because there is more than one cluster inferred for several of the main types (e.g., $2$ for both city and airport).  This is because there can be several files that seem to have come from the same source, and thus that are very similar to each other, but rather less similar to other files representing the same type in the ground truth.  We have not reported on the relationships because no foreign keys are detected between types; this reflects the heterogeneity of web table data.

In relation to the cultural data, fewer clusters are inferred (3) than appear in the ground truth (4).  This is because the inferred clusters put {\it Eurovision song contest entries} in the same cluster as {\it song}, which is reasonable.  However, although there is a cluster of films and of musical works, quite a few individual films and musical works tables are not included in any clusters, giving an overall rand score of {\it 0.62}. This reflects the fact that tables with quite diverse semantics can be associated with the same concept in the ground truth.  For example, an annotation of {\it film} can include rankings of best films, films for rent, and collections that are hard to interpret due to missing column names. As with the cities data, we have not reported on the relationships because no foreign keys are detected between different types; this is as expected for these data sets, where the different types are largely independent of each other.

\subsection{Runtimes}

The runtimes in Table \ref{tab:clustering} highlight the times taken by the different components of the approach. It is assumed, in operating on these data sets that are smaller than a complete data lake, that the approach is being applied to the results of an initial discovery process.  It can be seen that the times for clustering and relationship inference are both much smaller than those of the supporting tasks.  Note, however, that clustering may be repeated many times while tuning hyperparameters, whereas the other times take place only once for each application of the approach.

\section{Conclusions}
\label{sec:conclusions}

It is widely quoted that data scientists spend in the region of 80\% of their time preparing data for analysis, and within that as much as 20\% may be spent on collecting data sets\footnote{https://www.forbes.com/sites/gilpress/2016/03/23/data-preparation-most-time-consuming-least-enjoyable-data-science-task-survey-says/?sh=4ded67636f63}.  Data lakes hope to make relevant data more readily available. However, the schema-on-read model, in which the schema to be used for analysis is populated immediately prior to the analysis, defers the pain of selecting and integrating relevant data sets to analysis time.  As such, it is important to provide effective tools for discovering and interpreting data within data lakes.  
This paper has presented an approach to schema inference for data lakes that reuses indexing structures that have been shown to be effective for dataset discovery.  In so doing, it has made progress with respect to the contributions from the introduction:

\begin{enumerate}
	\item {\it The identification of the potential for schema inference in data lakes. } 
In particular, we have developed a proposal that uses existing approximate indexes for dataset discovery to provide the evidence needed both to identify the entity types and to infer the relationships that constitute a conceptual model for data lakes.  We have shown how this complements existing approaches to navigating through relationships within data lakes, providing a model that is both generic and abstract.
	\item {\it The provision of a technique for inferring schemas for data lakes. } %
We have provided a proposal that applies density-based clustering to infer entity types that represent multiple data sets, and have developed a proposal that automatically identifies how relationships between data lake tables can be combined to provide relationships between inferred entity types.
	\item {\it The evaluation of the resulting proposal using real-world data repositories. }
These experiments have shown promising results both for inferring entity types and for merging relationships between sources to identify relationships between entity types.
\end{enumerate}

We see schema inference as a promising addition to techniques for data lakes, providing a representation that is compact, and that emphasizes common ground among data sets.  Thus, for example, inferred schemas can potentially be used alongside search, to make explicit the context within which search results are found.  This paper has provided a first proposal for schema inference for data lakes.  There are a number of directions for further development.  For example, these include inferring names for entity types and relationships, identifying cardinality constraints in relationships, supporting richer target schemas for example with {\it is-a} relationships, the use of different distance functions for clustering, and the identification of additional sources of evidence to inform the clustering such as ontologies.
As such, though in this paper we have set the ball rolling in relation to schema inference for data lakes, we observe that this is a rich and challenging area, and that there are significant opportunities to explore with a view to creating more comprehensive schema-based views over data lakes.


\bibliographystyle{abbrv}
\bibliography{bibliography}

\end{document}